# Spin-pump-induced spin transport in a thermally-evaporated pentacene film

Yasuo Tani,[1] Yoshio Teki,[2] and Eiji Shikoh[1,a)]

[1]*Graduate School of Engineering, Osaka City University, 3-3-138 Sugimoto, Sumiyoshi-ku, Osaka 558-8585, Japan*

[2]*Graduate School of Science, Osaka City University, 3-3-138 Sugimoto, Sumiyoshi-ku, Osaka 558-8585, Japan*

We report the spin-pump-induced spin transport properties of a pentacene film prepared by thermal evaporation. In a palladium(Pd)/pentacene/$Ni_{80}Fe_{20}$ tri-layer sample, a pure spin-current is generated in the pentacene layer by the spin-pumping of $Ni_{80}Fe_{20}$, which is independent of the conductance mismatch problem in spin injection. The spin current is absorbed into the Pd layer, converted into a charge current with the inverse spin-Hall effect in Pd, and detected as an electromotive force. This is clear evidence for the pure spin current at room temperature in pentacene films prepared by thermal evaporation.

[a]E-mail: shikoh@elec.eng.osaka-cu.ac.jp

Pure spin current is dissipation-less information propagation in electronic devices. Carbon-based molecules are promising materials from the viewpoint of such spin transport because their spin-orbit interaction functioning as spin scattering centers is generally weak. In the field of molecular spintronics,[1-10] pure spin current has been observed in graphene[2,3] and organic polymers,[4,10] although the samples in those two cases were fabricated by the Scotch tape method and spin-coating, respectively, which are unsuitable for use in device fabrications.

In this study, we focus on a pentacene molecular film prepared by thermal evaporation (hereafter, "evaporated-pentacene" film) as a candidate material for spin transport. The reasons are as follows: First, vacuum thermal evaporation is better suited to preparing molecular films for electronic devices than spin-coating or the transfer of organic molecules. Secondly, pentacene has good crystallinity, even in films formed by thermal evaporation, and shows relatively high electrical conductivity without any dopants.[11] Pentacene, which is a *p*-type semiconductor, is well known as an active-layer material in organic field-effect transistors (OFETs).[11] The carrier mobility for pentacene film is above 1 $cm^2$/Vs, which is the highest value for any OFET with molecular films fabricated by thermal evaporation and high enough for simple electronic devices.[11] This means that numerous active carriers are present in the pentacene film (in this case, holes) to propagate the spin angular momentum. Moreover, pentacene shows photo-conductivity for visible light,[12] where the spin transport properties of pentacene can be controlled through light irradiation. This is an advantage for the use of molecular materials in future applications of spintronics over the use of inorganic materials. To date, spin transport in pentacene films has been studied using a spin-polarized charge current, as distinct from pure spin current, has been studied,[5,6] although there is a conductance mismatch problem[13,14] between the ferromagnetic electrode as a spin injector and pentacene. This problem lowers the spin injection efficiency and produces errors in the estimation of spin transport properties. In the present study, we show clear evidence for spin transport in a pentacene film at room temperature (RT) by using a pure spin current induced by spin-pumping.[15,16] In this case, the conductance mismatch problem related to spin injection with spin-pumping is negligible.[17-19]

Our sample structure and experimental set up is illustrated in Figure 1. Spin transport in pentacene is observed as follows: in palladium(Pd)/evaporated-pentacene/Ni$_{80}$Fe$_{20}$ tri-layer samples, a spin-pump-induced pure spin current, $J_S$, driven by ferromagnetic resonance (FMR) [15,16] of the Ni$_{80}$Fe$_{20}$ film is generated in the pentacene layer. This $J_S$ is then absorbed into the Pd layer. The absorbed $J_S$ is converted into a charge current as a result of the inverse spin-Hall effect (ISHE) [20] in Pd and detected as an electromotive force, $E$,[17-21] which is expressed as,

$$\vec{E} \propto \theta_{SHE} \vec{J}_S \times \vec{\sigma} \quad , \tag{1}$$

where $\theta_{SHE}$ is the spin-Hall angle, which is the efficiency of conversion from a spin current to a charge current, and $\sigma$ is the spin-polarization vector in the $J_S$. That is, if electromotive force due to the ISHE in Pd is detected under the FMR of $Ni_{80}Fe_{20}$, it is clear evidence for spin transport in a pentacene film.

Electron beam (EB) deposition was used to deposit Pd (Furuuchi Chemical Co., Ltd., 99.99% purity) to a thickness of 10 nm on a thermally-oxidized silicon substrate, under a vacuum pressure of $<10^{-6}$ Pa. Next, also under a vacuum pressure of $<10^{-6}$ Pa, pentacene molecules (Sigma Aldrich Co., Ltd.; sublimed grade, 99.9%) were thermally evaporated through a shadow mask. The deposition rate and the substrate temperature during pentacene depositions were set to 0.1 nm/s and RT, respectively, similar conditions under which pentacene films show high crystallinity.[22] The pentacene layer thickness, $d$, was varied between 0 and 100 nm. Finally, $Ni_{80}Fe_{20}$ (Kojundo Chemical Lab. Co., Ltd., 99.99%) was deposited by EB deposition through another shadow mask, under a vacuum pressure of $<10^{-6}$ Pa. During $Ni_{80}Fe_{20}$ deposition, the sample substrate was cooled with a cooling medium of -2°C, to prevent the deposited molecular films from breaking. As a control experiment, samples with a Cu layer, instead of the Pd layer, were prepared.

We used a microwave $TE_{011}$-mode cavity in an electron spin resonance (ESR) system (JEOL, JES-TE300) to excite FMR in $Ni_{80}Fe_{20}$, and a nano-voltmeter (Keithley Instruments, 2182A) to detect electromotive forces from the samples. Leading wires for detecting the output voltage properties were directly attached with silver paste at both ends of the Pd (or Cu) layer.

All of the measurements were performed at RT.

Figure 2(a) shows the FMR spectrum of a sample with a Pd layer and with the $d$ of 50 nm at an external magnetic field orientation angle $\theta$ of 0°, under a microwave power of 200 mW. The FMR field ($H_{FMR}$) of the $Ni_{80}Fe_{20}$ film is 120 mT and the $4\pi M_s$ of the $Ni_{80}Fe_{20}$, where $M_s$ is the saturation magnetization of the $Ni_{80}Fe_{20}$ film, is estimated to be 729 mT at a microwave frequency $f$ of 9.45 GHz and under FMR conditions in the in-plane field:

$$\frac{\omega}{\gamma} = \sqrt{H_{FMR}(H_{FMR} + 4\pi M_S)}, \qquad (2)$$

where $\omega$ and $\gamma$ are the angular frequency $2\pi f$ and the gyromagnetic ratio of $1.86\times10^{11}$ T$^{-1}$s$^{-1}$ of $Ni_{80}Fe_{20}$, respectively.[21] Fig. 2(b) shows the output voltage properties of the same sample shown in Fig. 2(a); the circles represent experimental data and the solid lines are the curve fit obtained using the equation[17-21]:

$$V(H) = V_{ISHE}\frac{\Gamma^2}{(H-H_{FMR})^2+\Gamma^2} + V_{Asym}\frac{-2\Gamma(H-H_{FMR})}{(H-H_{FMR})^2+\Gamma^2}, \qquad (3)$$

where $\Gamma$ denotes the damping constant (11.3 mT in this study). The first and second terms in eq. (3) correspond to the symmetry term to $H$ corresponding to the ISHE and the asymmetry term to $H$ (e.g. anomalous Hall effect [17-21] and other effects showing the same asymmetric voltage behavior relative to the $H$, like parasitic capacitances), respectively. $V_{ISHE}$ and $V_{Asym}$ correspond to the coefficients of the first and second terms in eq. (3). Output voltages are observed at $H_{FMR}$ at $\theta$ of 0 and 180°. Notably, the output voltage changes sign between $\theta$ values of 0 and 180°.

This sign inversion of voltage in Pd associated with the magnetization reversal in $Ni_{80}Fe_{20}$ is characteristic of ISHE.[17,19,21]

As a control experiment, we tested samples with a Cu layer, where the spin-orbit interaction is relatively weak, instead of the Pd layer. Figure 3(a) shows the FMR spectrum of a sample with a Cu layer and with the $d$ of 50 nm at the $\theta$ of 0°, under a microwave power of 200 mW. Fig. 3(b) shows output voltage properties for the same sample as in Fig. 3(a), where no clear electromotive force was observed at $\theta$ values of 0 and 180°. As another control experiment, we studied the microwave power ($P$) dependence of the electromotive forces; the results are shown in Fig. 4. The $V_{ISHE}$ increases in proportion to the increase in $P$. The above results suggest that the electromotive forces at the FMR field ($H - H_{FMR} = 0$) observed for the sample with a Pd layer (see Fig. 2(b)) are mainly due to the ISHE in Pd, that is, spin-pump-induced spin transport is achieved in an evaporated pentacene film at RT.

Figure 5 shows the $d$ dependence of (a) $4\pi M_s$ calculated via eq. (2) and of (b) $V_{ISHE}$ estimated via eq. (3). With increasing $d$, $M_s$ decreases slightly, while $V_{ISHE}$ clearly decreases. The data for the pentacene-free case ($d = 0$ nm) are plotted as open circles in Fig. 5(b). $V_{ISHE}$ in the pentacene-free case, however, includes extrinsic effects, e.g., the electromotive force due to $Ni_{80}Fe_{20}$ itself.[23] Thus, using all data except those for $d = 0$ nm, we estimated the spin diffusion length of the evaporated-pentacene film $\lambda_s$ to be 42±10 nm at RT under the assumption[4,17] of an exponential decay of the spin current in the pentacene film. The dashed lines in Fig. 5 represent the results of this estimation. To validate these estimates, we also estimated $\lambda_s$ in another way,[17] using the linewidth of the FMR spectra, the $4\pi M_s$ calculated via eq. (2), and the experimentally obtained electromotive forces results. The detailed calculation method is described in ref. 17. The real part of the mixing conductance at the interface between $Ni_{80}Fe_{20}$ and the Pd layer, $g_r^{\uparrow\downarrow}$, which could be thought of as the transmittance of the spin current, and the generated spin current density in the Pd layer, $j_s$', are calculated to be $2.52\times10^{19}$ m$^{-2}$ and $7.31\times10^{-10}$ Jm$^{-2}$, respectively. This $g_r^{\uparrow\downarrow}$ is almost the same as that obtained in experiments involving spin-pump-induced spin transport in $p$-type Si.[17] Meanwhile, the $j_s$' in the present study is by two orders of magnitude smaller than in the case of the $p$-Si.[17] Using the spin diffusion length of Pd, 9 nm [24] and the $\theta_{SHE}$ of Pd, 0.01,[21] the $\lambda_s$ is estimated to be about 38.5 nm. The above two $\lambda_s$ estimates are almost the same, although the $\lambda_s$ value of 30-40 nm at RT is comparable to or shorter than the spin diffusion lengths of other molecules: ~150 nm for a π-conjugated polymer (PBTTT),[4] ~150 nm for a $p$-type conducting polymer (PEDOT:PSS),[10] and ~50 nm for a low-molecular-weight $n$-type molecule (Alq$_3$).[8,9] Pentacene is a low-molecular-weight $p$-type semiconductor. Thus, the polarity of the major carrier of the molecules is not strongly affected by the shortening of the spin diffusion length. If a detailed investigation of the relationship between the carrier polarity and the spin transport mechanism is necessary, a study of spin transport using ambipolar molecules would be effective. Polymers, such as PBTTT and PEDOT:PSS, tend to possess longer spin diffusion lengths than low-molecular-weight molecules, such as Alq$_3$ and pentacene. We checked crystallinity of our pentacene film grown on the Pd layer by using an X-ray diffractometer (See supplemental material [25]). In our pentacene films, the pentacene molecules might partially be oriented. In general, Alq$_3$ molecular film has amorphous state. Comparing pentacene with Alq$_3$, the crystallinity of molecular films may be related to the spin diffusion length. That is, the higher the crystallinity of a molecular film is, the longer the spin diffusion length of the molecular film might be. Thus, to improve the crystallinity of pentacene films is an indispensable issue to clarify the mechanism of spin transport in a pentacene film. The mechanism of spin transport among the above molecules may be different because the charge transport mechanism is different for each. At present, the reason why polymers show longer spin diffusion lengths is still unclear, except for the crystallinity of molecular films. More studies of the spin transport in various molecules having various charge transport mechanisms and having high crystallinity are eagerly awaited to clarify the spin-transport mechanism in molecular films.

In summary, spin transport properties of evaporated-pentacene films were studied at RT by using spin-pumping for spin injection and using the ISHE in non-magnetic metals in the spin detection methods. We achieved spin transport in evaporated-pentacene films; the spin diffusion length in pentacene was estimated to be above 30 nm at RT. This is clear evidence for pure spin current in molecular films prepared by thermal evaporation, which paves the way to molecule-based spintronic devices.

The authors thank to Mr. Yutaka Ohkawahara and Prof. Dr. Masaaki Nakayama in Osaka City Univ. Japan, for helping to evaluate the crystallinity of pentacene films. This research was partly supported by a Grant-in-Aid from the Japan Society for the Promotion of Science (JSPS) for Scientific Research (B) (26286039 (to E. S.)), a Grant-in-Aid from the Japan Society for the Promotion of Science (JSPS) for Scientific Research (B) (24350076 (to Y. Te.)) and Jyuten Kenkyu (B) from Osaka City University.

Figure captions

FIG. 1. (a) Bird's-eye-view and (b) top-view illustrations of our sample and orientations of external applied magnetic field $H$ used in the experiments. $J_S$ and $E$ correspond, respectively, to the spin current generated in the pentacene film by spin-pumping and the electromotive forces due to the ISHE in Pd.

FIG. 2. (a) FMR spectrum and (b) output voltage properties of a sample with a Pd layer and with a pentacene film thickness $d$ of 50 nm, under a microwave power of 200 mW.

FIG. 3. (a) FMR spectrum and (b) output voltage properties of a sample with a Cu layer and with a pentacene film thickness $d$ of 50 nm, under a microwave power of 200 mW.

FIG. 4. (a) Microwave power ($P$) dependence of electromotive force and (b) analysis results obtained with eq. (2). $V_{ISHE}$ and $V_{Asym}$ correspond to the coefficients of the first and second terms in eq. (2).

FIG. 5. Dependence of (a) $4\pi M_s$ ($M_s$: saturation magnetization), calculated by eq. (1), and of (b) $V_{\text{ISHE}}$ estimated by eq. (2), on the pentacene film thickness ($d$). Open and closed circles in (b) are the experimental data. The dashed lines in (b) are curve fits under the assumption of exponential decay.

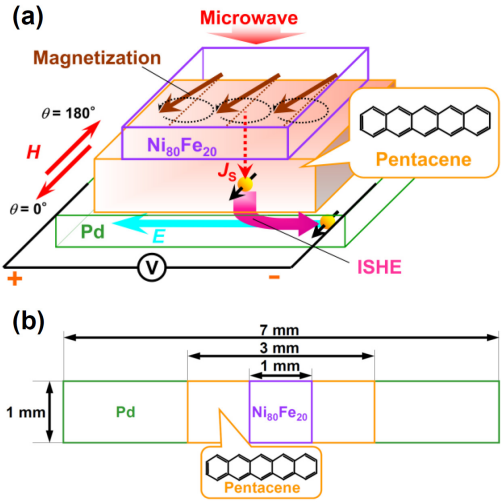

Y. Tani, et al.:   FIG. 1.

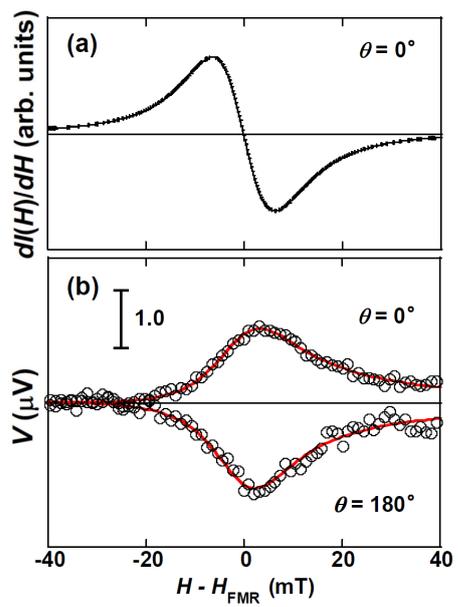

Y. Tani, et al.:   FIG. 2.

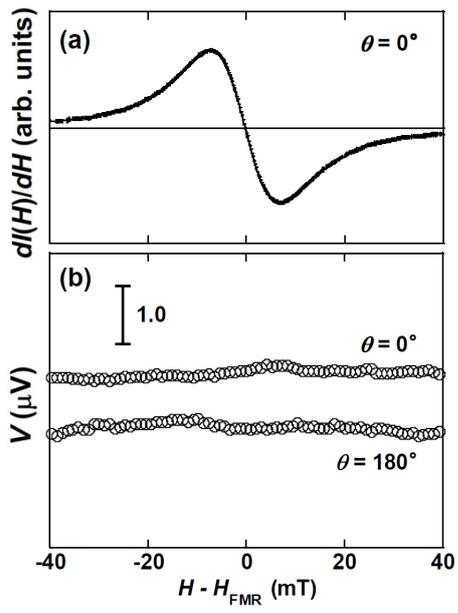

Y. Tani, et al.: FIG. 3.

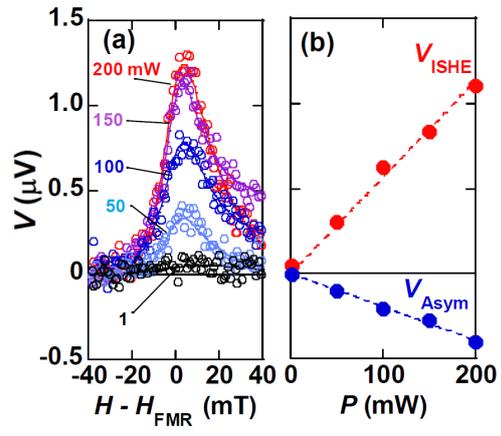

Y. Tani, et al.: FIG. 4.

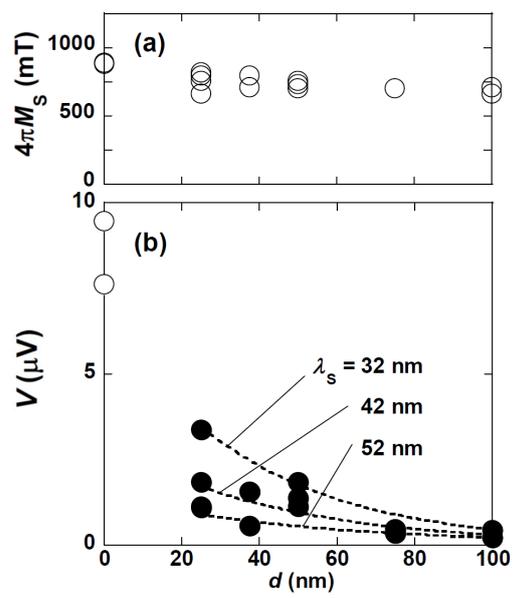

Y. Tani, et al.:   FIG. 5.

Supplemental material for:

Spin-pump-induced spin transport in a thermally-evaporated pentacene film

by Yasuo Tani,[1] Yoshio Teki,[2] and Eiji Shikoh[1]

[1]*Graduate School of Engineering, Osaka City University, Japan*

[2]*Graduate School of Science, Osaka City University, Japan*

The crystallinity of our pentacene film was checked by an X-ray diffractometer (Shimadzu Corp., XRD-6000). A general $\theta$-$2\theta$ scan method was used. FIG. S1 shows the XRD spectrum of a pentacene film grown on a Pd layer. The thicknesses of the pentacene film and Pd layer are 100 nm and 10 nm, respectively. The X-Ray wavelength was 0.154 nm (Cu-K$\alpha$). The diffraction peaks were identified by using a reference paper.[S1] The (001) and (002) diffraction peaks corresponding to the perpendicular orientation to the film plane were observed. The XRD measurement suggest that the pentacene molecules in our film might partially be oriented.

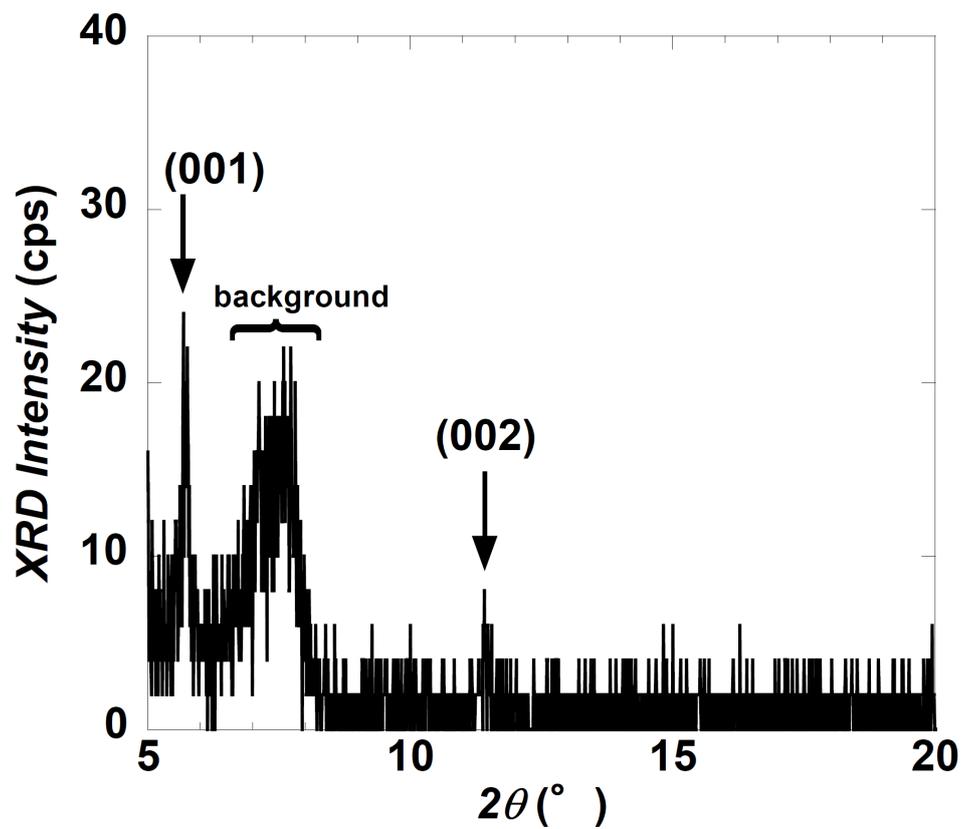

FIG. S1. XRD spectrum of a pentacene film grown on a Pd layer. The diffraction peaks are identified by using a reference paper.[S1]

Reference

[S1] C.D. Dimitrakopoulos and P.R.L. Malenfant, Adv. Mater. **14**, 99 (2002).